\documentclass[twocolumn,prl,superscriptaddress,showpacs]{revtex4-1}
\usepackage{amsmath,amssymb}
\usepackage{graphicx}
\usepackage{verbatim}
\usepackage{amsfonts}
\usepackage{dcolumn}
\usepackage{bm}
\usepackage{color}
\usepackage[colorlinks,citecolor=blue]{hyperref}

\begin{document}

\title{\large \bf Microscopic Realization of 2-Dimensional Bosonic Topological Insulators}

\author{Zheng-Xin Liu}
\affiliation{Institute for Advanced Study, Tsinghua University, Beijing, 100084, P. R. China}
\affiliation{Perimeter Institute for Theoretical Physics, Waterloo, Ontario, N2L 2Y5 Canada}
\author{Zheng-Cheng Gu}
\affiliation{Perimeter Institute for Theoretical Physics, Waterloo, Ontario, N2L 2Y5 Canada}
\author{Xiao-Gang Wen}
\affiliation{Institute for Advanced Study, Tsinghua University, Beijing, 100084, P. R. China}
\affiliation{Perimeter Institute for Theoretical Physics, Waterloo, Ontario, N2L 2Y5 Canada}
\affiliation{Department of Physics, Massachusetts Institute of Technology, Cambridge, Massachusetts 02139, USA}

\begin{abstract}

It is well known that a Bosonic Mott insulator can be realized by condensing vortices
of a boson condensate. Usually, a vortex becomes an anti-vortex (and vice-versa)
under time reversal symmetry, and the condensation of vortices results in a trivial Mott
insulator. However, if each vortex/anti-vortex interacts with a spin trapped at its core, the time reversal transformation of the composite vortex operator will contain an extra minus sign. It turns out that such a composite vortex condensed state is a bosonic topological insulator (BTI) with gapless boundary excitations protected by $U(1)\rtimes Z_2^T$ symmetry.  We point out that in BTI, an external $\pi$ flux monodromy defect carries a Kramers doublet. 
We propose lattice model Hamiltonians to realize the BTI phase, which might be implemented in cold atom systems or spin-$1$ solid state systems.

\end{abstract}

\maketitle

\textit{Introduction.}
Quantum phases beyond Landau symmetry breaking theory\cite{L3726,GL5064,LanL58}, 
including long-range entangled\cite{CGW1038} intrinsically topologically ordered
states\cite{Wtop,WNtop,Wrig} and short-range entangled symmetry-protected
topological (SPT) states\cite{GW0931,PBT0959,CLW1141,CGL1314,CGL1204}, have attracted great interest in condensed matter physics recently. 
Intrinsic topologically ordered states, such as fractional quantum
Hall states, can be characterized by their bulk fractionalized
excitations. On the other hand, SPT phases do not have nontrivial bulk excitations and can be adiabatically connected to
a trivial product state if symmetry is broken in the bulk. The 1D Haldane phase for spin-1 chain\cite{H8364,AKL8877,GW0931,PBT0959} and
topological insulators\cite{KM0501,BZ0602,KM0502,MB0706,FKM0703,QHZ0824} are
non-trivial examples of SPT phases. Bosonic SPT
phases in spacial dimension $d$ are (partially) classified by the $(d+1)$th
group cohomology of the symmetry group \cite{CGL1314}. Some SPT phases can also
be described by Chern-Simons theory (2D) \cite{LV1256,LW1224, CG13} or topological terms of non-linear sigma model\cite{BRX1315}. Typically, nontrivial SPT phases can be characterized by their edge excitations. In 1D, the edge state of a SPT phase is
degenerate and transform as projective representation of the symmetry group
\cite{CGW1107,CGW1128}. In 2D, the edge state of SPT phase is either gapless or
symmetry breaking \cite{CLW1141,CGL1314}. The boundary theory of 3D SPT phases
is more interesting, it can be either gapless \cite{YeWang13} or topologically ordered if
symmetry is unbroken\cite{VS1306,MKF1335}. 
It is worthwhile to mention that if the
ground state is long-range entangled, symmetry can also act on the bulk topological
excitations nontrivially, resulting in different symmetry enriched topological (SET)
phases\cite{W0213,KLW0834,KW0906,MR1315,HW1351,LV1334}.

\begin{figure}[t]
\centering
\includegraphics[width=3.4in]{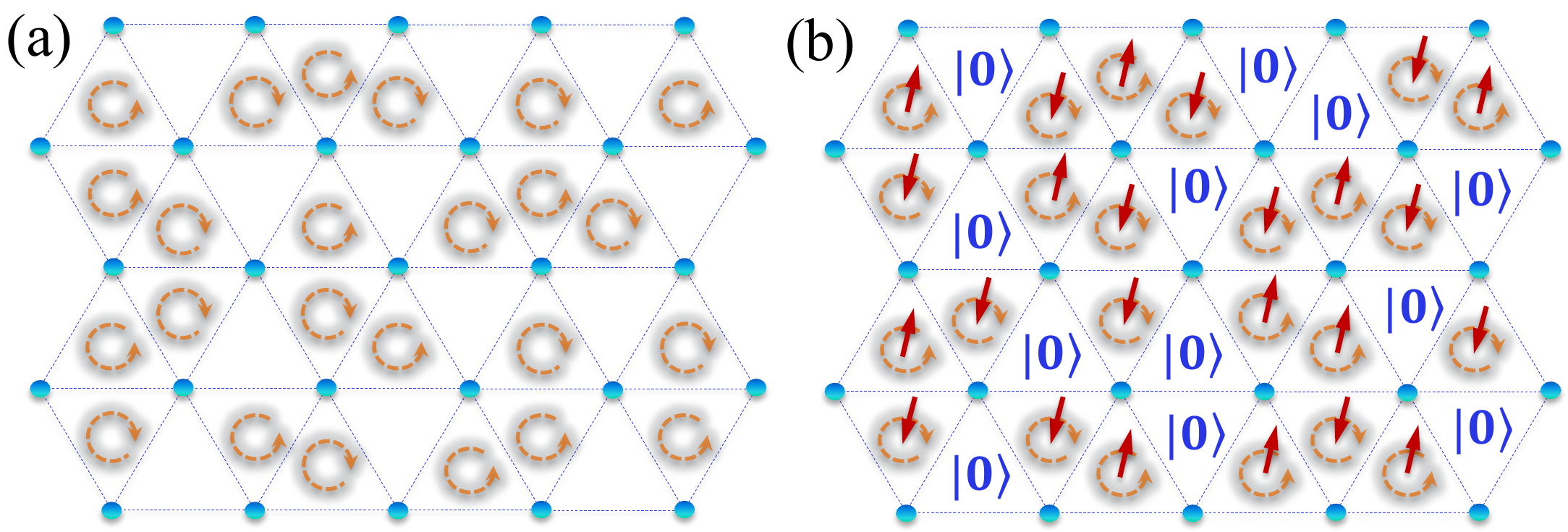}
\caption{(Color online) Two kinds of bosonic Mott insulators. The dots are bosons and the circles means vortex/anti-vortex. (a) Condensing the vortex/anti-vortex results in a trivial bose Mott insulator; 
(b) 
If $S=1$ spins exist on the dual lattice, and creating a vortex/anti-vortex creation flips the spin momentum of the spin at its core, then after condensing the composite vortex/anti-vortex the resultant state is a bose topological Mott insulator. Here $|0\rangle$ means $S_z|0\rangle=0$, and uparrows and downarrows means $|1\rangle$ and $|-1\rangle$ respectively. } \label{fig:BTI}
\end{figure}

In analogy to fermionic topological insulators, Bosonic SPT phases protected by $U(1)\rtimes Z_2^T$ symmetry (where $U(1)=\{U_\theta; U_\theta=e^{i\theta}\}$, $Z_2^T=\{I, T\}$ and $U_\theta T=TU_{-\theta}$) are called bosonic topological (Mott) insulators. One kind of nontrivial 3D Bosonic topological insulators have been discussed via quantum field theory approach\cite{VS1306,MKF1335} and projective construction\cite{YW1372}. In this paper, we will discuss how to realize 2D bosonic topological insulators (BTI). Since $\mathcal H^3(U(1)\rtimes Z_2^T,U(1))=\mathbb Z_2$, there are two Mott insulating phases, one is trivial and the other is the nontrivial BTI.
Our construction of the SPT phases contains two steps. The first step is condensing the boson field to break the $U(1)$ symmetry. The second step is condensing the vortex field to restore the broken symmetry. In the usual case, the vortex is charge neutral and changes into anti-vortex under time reversal. The resultant state after vortex condensation is a trivial Mott insulator. However, if there exists another layer of spins living on the dual lattice such that each vortex core traps a spin \cite{WSC9794,YTQW2011} (see Fig. \ref{fig:BTI}b), the sign of the composite vortex operator is reversed under time reversal owing to interaction. The resultant state is a nontrivial BTI\cite{XS1372}, whose gapless edge states are symmetry protected. We illustrate that in the BTI a monodromy defect ({\it i.e.} a $\pi$-flux) traps a Kramers doublet. Finally, we discuss Hamiltonian 
realization of the BTI in lattice models. These lattice Hamiltonians may shed light on further numerical studies and experimental realization of more nontrivial SPT phases.

\textit{Bosonic Mott insulators with $U(1)\rtimes Z_2^T$ symmetry.}  Both the trivial and nontrivial bosonic Mott insulators can be constructed from Bose condensate. The idea is first condensing the bosons to break the $U(1)$ symmetry and then restoring the symmetry by condensing vortices.

Let us start with the trivial bosonic Mott insulator. Consider a familiar contact interacting boson model respecting $U(1)\rtimes Z_2^T$ symmetry:
\begin{eqnarray}
\mathcal L_{ b}={1\over2}( b^*\partial_\tau b + h.c) + {1\over2m}\partial_x b^*\partial_x b - \mu| b|^2 + {g\over2}| b|^4,
\end{eqnarray}
when $\mu>0, g>0$, the bose field $ b(x)$ condenses to a classical value $\sqrt{\rho_0}=\sqrt{\mu\over g}$  and the $U(1)$ symmetry is spontaneously broken. We can write $ b(x)=\sqrt{\rho_0+\delta\rho} e^{i\theta(x)}$, where $\delta\rho$ and $\theta(x)$ are the density and phase fluctuations of the condensation, respectively. Integrating out the magnitude fluctuation $\delta \rho$, we obtain an XY model as the low energy effective Lagrangian\cite{Wen04}:
\begin{eqnarray}\label{XY1}
\mathcal L_{\rm XY} = {1\over2g} [(\partial_\tau \theta)^2 + v^2(\nabla\theta)^2],
\end{eqnarray}
where $v=\sqrt{\rho_0g\over m}$ is the sound velocity. The boson density and current fluctuations are given by $J_0\equiv\rho=\partial_\tau \theta$ and $J_i=\rho_0\partial_i\theta$, respectively.

The XY model (\ref{XY1}) is obviously gapless. There are two ways to open a gap. One way is to add a term $G\cos\theta$ to the model. This term explicitly breaks the $U(1)$ symmetry. The other way is to condense the vortex field. Condensing the vortex field will restore the $U(1)$ symmetry and the resultant state is a Mott insulating phase. We can introduce another bose field $\tilde b$ to describe the vortex condensation. As derived in the supplemental material, the bose field $b$ and the vortex field $\tilde b$ can be described by two gauge fields $a$ and $\tilde a$ respectively, where the boson current is given by $J^\mu={i\over2\pi}\varepsilon^{\mu\nu\lambda}\partial_\nu a_\lambda$ and the vortex current is given by  $\tilde J^\mu={i\over2\pi}\varepsilon^{\mu\nu\lambda}\partial_\nu \tilde a_\lambda$. The two gauge fields have a mutual Chern-Simons coupling and the effective Lagrangian is
\begin{eqnarray}\label{doubleCS}
\mathcal L_{\mathrm{CS}}
&=&-{i\over4\pi}K_{IJ}\varepsilon^{\mu\nu\lambda}a_{I\mu}\partial_\nu a_{J\lambda} + \mathcal L_{\mathrm{Maxwell}},
\end{eqnarray}
where $K=\left(\begin{matrix}0&1\\1&0\end{matrix}\right)$ and $I,J=1,2$. Here have we changed the notation $a_{1\mu}=\tilde a_\mu$, $a_{2\mu}=a_\mu$ and the corresponding boson fields $ b_1= b$, $ b_2=\tilde b$. Under this notation, $b_1$ (the boson) carries charge of $a_{1\mu}$ gauge field and $b_2$ (the vortex) carries charge of $a_{2\mu}$ gauge field. The charges of both gauge fields $a_1, a_2$ are quantized since the boson number and vortex number are quantized. Therefore, the Lagrangian (\ref{doubleCS}) is a well defined effective field theory for the system.

If we gauge the $U(1)$ symmetry and probe the system through such an external gauge field $A_\mu$, a coupling term should be added to the Lagrangian (\ref{doubleCS}):
$\mathcal L_{\rm prob} ={i\over 2\pi}q_I\varepsilon^{\mu\nu\lambda}A_\mu\partial_\nu a_{I\lambda},$
where the charge vector is given by $q=(0,\ 1)^T$ \footnote{Noticing that $J_{2\mu}={1\over2\pi}\varepsilon_{\mu\nu\lambda}\partial_\nu a_{2\lambda}={1\over2\pi}\varepsilon_{\mu\nu\lambda}\partial_\nu a_{\lambda}$ is the boson current, it is natural that the charge vector is equal to $(0,\ 1)^T$.   This is consistent with the factor that the $b_1$ boson [described by particle vector $l=(1,\ 0)^T$] carries $U(1)$ symmetry charge, $Q=l^TK^{-1}q=1$.}. The response to the probe field $A_\mu$ is the Hall conductance $\sigma_{\rm H}={1\over2\pi}q^TK^{-1}q$. It is easy to see that above Mott insulator has zero Hall conductance (it is the same with the nontrivial bosonic topological insulator that will be discussed below) .

From the theory of quantum Hall effect\cite{W9125}, the boundary of (\ref{doubleCS}) is given by $\mathcal L_{\rm bdry}=-{i\over4\pi}K_{IJ}\partial_\tau\phi_I\partial_x\phi_J+V_{IJ}\partial_x\phi_I\partial_x\phi_J$, where $\partial_i\phi_I=a_{Ii}$, and the boson density operator $\rho_I$ and creation operator $b_I$ can be expressed in forms of the $\phi_I$ field as
\begin{eqnarray}\label{Conform}
\rho_I=\partial_x\phi_I,\ \ b_I\sim e^{-i\phi_I}.
\end{eqnarray}

Before further discussion, let us see how the boson fields change under the symmetry group. Recalling that the boson field $ b_1$ carries $U(1)$ symmetry charge and the vortex field $ b_2$ is neutral, and the vortex becomes an antivortex under time reversal, they vary in the following way under $U(1)\rtimes Z_2^T$ symmetry: $U_\theta  b_1 U_\theta^{-1} = e^{-i\theta} b_1,\ T  b_1 T^{-1} = b_1$ \footnote{Notice that an arbitrary phase factor $T  b_1 T^{-1} = e^{-i2\phi}b_1$ arises if we redefine the boson operator $b_1\to b_1e^{i\phi}$. So this phase factor is not important and can be fixed to 1.},  $U_\theta  b_2 U_\theta^{-1} =  b_2,\ T  b_2 T^{-1} = b_{2}^{\dag}$. From eqn.~(\ref{Conform}), above relations can also be written as
\begin{eqnarray}\label{phi1phi2}
&&U_\theta  \phi_1 U_\theta ^{-1}= \phi_1+\theta,\ T  \phi_1 T ^{-1}=-\phi_1,\nonumber\\
&&U_\theta  \phi_2 U_\theta ^{-1}=  \phi_2,\ \ \ \  T  \phi_2 T^{-1} = \phi_{2},
\end{eqnarray}

The bulk of the doubled Chern-Simons theory is gapped. What is interesting is the edge spectrum. Since the field $\phi_I$ satisfy the following Kac-Moody algebra,
\begin{eqnarray}\label{KacMoody}
[\partial_x\phi_I,\partial_y\phi_J]=2\pi iK_{IJ}^{-1}\partial_x\delta(x-y),
\end{eqnarray}
the edge excitations may be gapless.  As pointed out in Ref.~\onlinecite{LV1256}, a perturbing term $G\cos\phi_2$ with $G<0$ locates the $\phi_2$ field to its classical value $\phi_2=0$ and gaps out the boundary without breaking any symmetry [see eqn.(\ref{phi1phi2})]. That is to say, it is a trivial SPT phase.


In a Mott insulator, the boson number per site is an integer. It can be simply realized in a lattice Hamiltonian --- the Bose Hubbard model
\begin{eqnarray}\label{TrivialSPT}
H=-\sum_{\langle ij\rangle}t_{ij} b_i^\dag b_j +\sum_i [U(b_i^\dag b_i-1)^2+\mu b_i^\dag b_i].
\end{eqnarray}
If $U$ is larger than a critical value $U_c$, the boson will be localized and the boson number per site approaches to one, which forms a (trivial) Mott insulator.

Since the bose Hubbard model only realizes the trivial bose Mott insulator, it is mystical how to construct the nontrivial one. It turns out that it  can be realized as long as the vortex of the bose condensate varies nontrivially under time reversal \cite{LV1256}. To this end, we couple the bose current to a second layer of $S=1$ spins on the dual lattice (see Fig. \ref{fig:BTI}b, the spins are neutral under $U(1)$ symmetry). Since a vortex/anti-vortex carries nontrivial bose current, we assume that under certain interaction creating a vortex/anti-vortex will increase/decrease the spin angular momentum of the $S=1$ spin at the vortex/anti-vortex core. 
Thus we obtain a composite vortex $ b_2'$ as a combination of $ b_2$ and $S^-$,
\[
b_2'= b_2S^-.
\]
where $S^\pm=S^x\pm iS^y$.

Notice that the spin is charge neutral $U_\theta S^-U_\theta^{-1}=S^-$ and reverse its sign under time reversal $TS^-T^{-1}=-({S^-})^\dag=-S^+$, so for the composite vortex, we have $U_\theta(b_2')U_\theta^{-1}=b_2',\ \  T(b_2')T^{-1}=-(b_2')^\dag.$ When the new vortex field condense, the $U(1)$ symmetry of the bulk in the boson layer is restored and a gap is open. Repeating previous argument, we can formulate the effective theory as a doubled Chern-Simons theory (\ref{doubleCS}). However, what is different here is that the boundary remains gapless if the $U(1)\rtimes Z_2^T$ symmetry is unbroken. To see this, we rewrite the new vortex as $b_2'=e^{-i\phi_2'}$ at the boundary, $\phi_2'$ varies in the following way under the symmetry group:
\begin{eqnarray}
U_\theta \phi_2' U_\theta^{-1} = \phi_2',\ \ \ \  T \phi_2' T ^{-1}= \phi_2' + \pi.
\end{eqnarray}
 $\phi_1$ varies in the same way as given in (\ref{phi1phi2}). $\phi_1$ and $\phi_2'$ still satisfy Kac-Moody algebra $[\partial_x\phi_1,\partial_y\phi_2']=2\pi i\partial_x\delta(x-y)$. To gap out the boundary, either $U(1)$ or time reversal symmetry $Z_2^T$ should be broken explicitly or spontaneously. For instance, the perturbation term $G\cos(2\phi_2')$ is invariant under the symmetry group and can gap out the boundary, but there are two ground states $\phi_2'=0$ and $\phi_2'=\pi$. Time reversal $T\phi_2'T^{-1}=\phi_2'+\pi$ transforms one ground state into the other, so the $Z_2^T$ symmetry is spontaneously broken.

Noticing that $T(b_2')T^{-1}=-(b_2')^\dag$ , careful readers may ask why time reversal symmetry is not broken by condensing $b_2'=b_2S^-$. At first glance, it seems that $T$ is broken since $\langle b_2'\rangle$ is not invariant under $T$. However, since $b_2'$ carries charge of the gauge field $a_2$, it is not gauge invariant. In other words, the Lagrangian (\ref{doubleCS}) is invariant under the following gauge transformation
\[
b_2' \to b_2'e^{i\varphi},\ a_2\to a_2-\partial\varphi.
\]
The sign change of $b_2'$ under time reversal can be compensated by a gauge transformation $b_2'\to-b_2'$. When averaging all the gauge fluctuations, the vortex condensate has a zero expectation value $\langle b_2'\rangle=0$ \cite{Elitzur75}.  In this way, we conclude that time reversal symmetry in not broken in the ground state even when $b_2'$ is condensed. 

 The BTI may also be realized in superfluids or superconductors (in the molecule limit) with spin orbital coupling. Supposing that the boson carries a soft spin-1 momentum which are initially staying at $|0\rangle$ state, and the vortex/anti-vortex flips the spin momentum at its core to $|1\rangle$/$|-1\rangle$ owning to spin-orbital coupling, then a nontrivial BTI is obtained if the vortex/anti-vortex condense.

\textit{Symmetry-protected invariants and possible lattice model realization of the BTI.} The non-triviality of the BTI constructed above can also be illustrated by its symmetry-protected invariant\cite{W1375}. The invariant of the BTI is that a $\pi$-flux of the $U(1)$ symmetry\footnote{The physical excitations of the system only contain integer times of $2\pi$ fluxes (namely, the vortices), so the $\pi$ flux can be seen as external perturbation, the so-called monodromy defect.} carries Kramers doublet. We will show how this is true.

In continuous field theory, the flux is dispersed in space and consequently a $\pi$-flux (noted as $|\pi\rangle$) is distinct from $-\pi$ flux (noted as $|-\pi\rangle$). Under time reversal, they transform into each other
\[
T|\pi\rangle\propto |-\pi\rangle,\  \ T|-\pi\rangle\propto |\pi\rangle.
\]
To construct a time reversal invariant $\pi$-flux centered at $x$, we have to consider a superposition of states $|\pi\rangle$ and $|-\pi\rangle$. On the other hand, $|-\pi\rangle$ and $|\pi\rangle$ are related by a vortex creation or annihilation operator,
\begin{eqnarray}\label{mvortex}
b_2'|\pi\rangle=\eta|-\pi\rangle,\ \ b_2'^\dag|-\pi\rangle=\eta |\pi\rangle,
\end{eqnarray}
where $\eta$ is a constant.
Notice that $Tb_2'^\dag T^{-1}=-b_2'$, we have $Tb_2'^\dag T^{-1}T |-\pi\rangle=\eta T|\pi\rangle=-b_2'T|-\pi\rangle$ and similarly $Tb_2' T^{-1}T |\pi\rangle= \eta T|-\pi\rangle=-b_2'^\dag T|\pi\rangle$. Comparing with (\ref{mvortex}), we can choose a proper gauge such that
\begin{eqnarray}\label{doublet}
&&T|\pi\rangle=|-\pi\rangle,\nonumber\\
&&T|-\pi\rangle=-|\pi\rangle.
\end{eqnarray}
This means that  when acting on the two-dimensional space $|\pi\rangle$ and $|-\pi\rangle$, time reversal $T$ behaves as $\hat T=i\sigma_yK$ satisfying $\hat T^2=-1$. Since $\hat T$ is irreducible, $|\pi\rangle$ and $|-\pi\rangle$ (generally $|(2n+1)\pi\rangle$ and $|-(2n+1)\pi\rangle$) form a Kramers' doublet. As a contrast, in the trivial Mott insulating phase, the time-reversal invariant $\pi$-flux is generally non-degenerate.

\begin{figure}[t]
\centering
\includegraphics[width=3.5in]{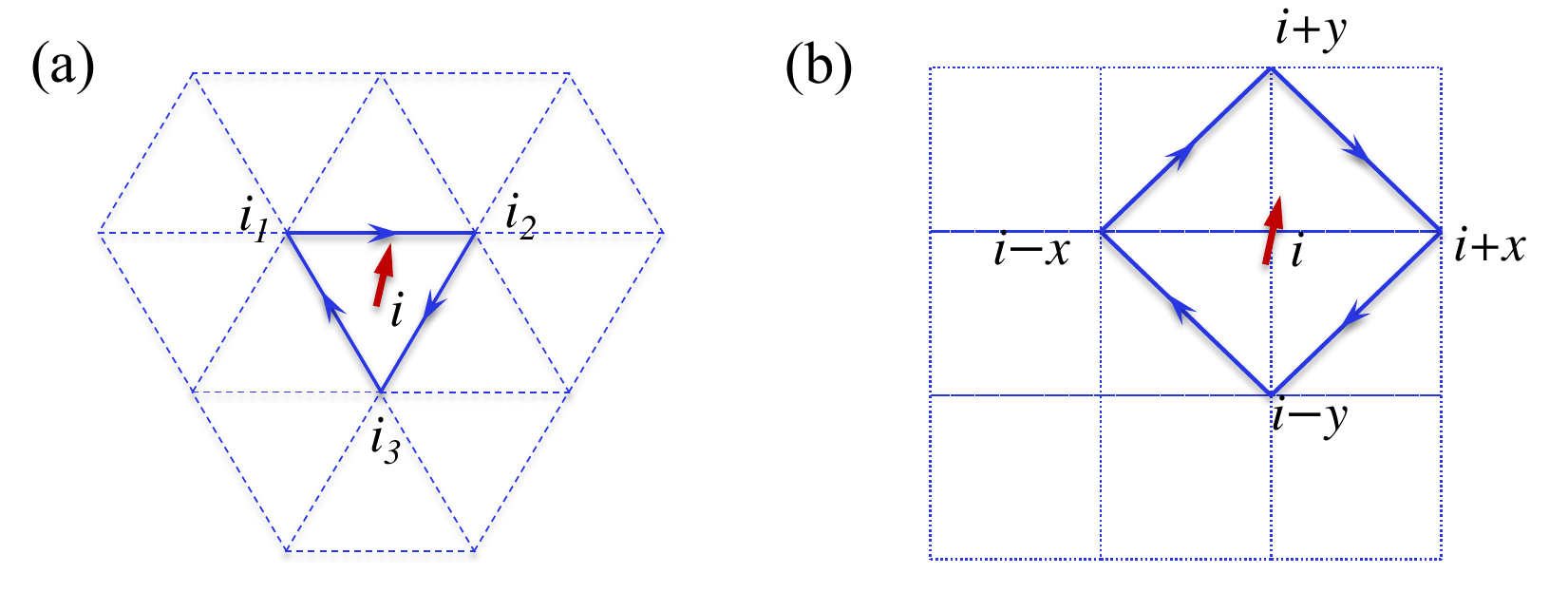}
\caption{(Color online) Interactions that stabilizes nontrivial SPT states. The bose current interact with the spin such that the spin feel an effective `Zeeman field' (coupling to $S_z$) induced by the vortex current. (a) BTI on a double layer system. The spinless bosons live on triangular lattice and the spins (at boson vortex cores) locates on the dual lattice;  (b) BTI of spin-1 bosons on square lattice. The vortex core of condensate on A sub-lattice locates on B sub-lattice, and vice versa.
} \label{fig:Int}
\end{figure}

In a lattice model, we don't need to distinguish $\pi$-flux and $-\pi$-flux. In this case, a $\pi$-flux is still a doublet since the degeneracy can never be removed unless time reversal symmetry is broken. This is the symmetry-protected invariant of the topological Mott phase and can be used to distinguish from the trivial one.

Above we have proposed a mechanism to obtain the nontrivial bose insulating state. Now we propose a possible interaction that may stabilize this state. The key point is to flip a spin when creating a vortex/anti-vortex. However, the vortex/anti-vortex creation operator is usually nonlocal and it is difficult to give its explicit form. Noticing that a vortex/anti-vortex carries bose current, we can couple the bose current with the spins. In addition to the Hamiltonian (\ref{TrivialSPT}), a possible $U(1)\rtimes Z_2^T$ invariant interaction is (see Fig.\ref{fig:Int}a)
\begin{eqnarray} \label{int1}
H_{\rm int}=-\sum_i g S_{i}^z(J_{i_1i_2} + J_{i_2i_3} + J_{i_3i_1}),
\end{eqnarray}
where $g>0$ is the interaction strength and $J_{ij}=(-ib_i^\dag b_j+h.c.)$ is the bose current operator. Here we consider triangular lattice for the boson, and the spins live on the dual Honeycomb lattice. The spins are initially staying in eigenstate of $S^z$, $S^z|0\rangle=0$ (Notice that the spin layer has time reversal symmetry but no spin rotational symmetry. The spins may have the following interaction $\sum_{\langle i,j\rangle} J\mathbf S_i\cdot\mathbf S_j +\sum_iD(S_i^z)^2$ \footnote{The Heisenberg interaction may explain that the $\pi$ flux monodromy defects, which can be considered as ends of a string, carry spin-1/2 Kramers doublets.} plus some perturbations to remove accidental symmetries). 
 If $g>D>J$, the spin at the vortex (or anti-vortex) core will be flipped from $|0\rangle$ to $|1\rangle$ (or $|-1\rangle$), in agreement with our expectation that $b_2'^\dag=b_2^\dag S^+$, $b_2'=b_2S^-$.

\textit{Realization of BTI in a single layer of spin-1 bosonic system. } In above discussion we proposed to realize the BTI in a double layer model. In the following we show that it can also be realized in a single layer of spin-1 bosons, which is more likely to be implemented in cold atom systems\cite{FeiBEC, LLC1D}.
A spin-1 boson can be considered as three species of bosons $b_\alpha$ ($\alpha=1,0,-1$). The $U(1)$ symmetry is defined by $U(1)=\{e^{i\hat N\theta}; \theta\in[0,2\pi)\}$, where $\hat N=\sum_{i}\hat N_i$,  $\hat N_i=\sum_{\alpha}b_{\alpha,i}^\dag b_{\alpha,i}$.  As usual, time reversal operator $T$ will reverse the spin and is defined as $T=e^{i\sum_iS_i^y\pi}K$, where $S_i^y=\frac{1}{\sqrt{2}}[-i(b_{1,i}^\dag b_{0,i}+b_{0,i}^\dag b_{-1,i})+h.c.]$ ($S_i^x, S_i^z$ are defined similarly) and $K$ means complex conjugation. When acting on the boson operators, we obtain $U_\theta b_\alpha U_\theta^\dag=b_\alpha e^{i\theta}$ ($U_\theta\equiv e^{i\hat N\theta}$), $Tb_0T^{-1}=-b_0, Tb_1T^{-1}=b_{-1}, Tb_{-1}T^{-1}=b_1$, and $TS_i^mT^{-1}=-S_i^m$ where $m=x,y,z$.

We consider square lattice, which contain two sub-lattices labeled as $A$ and $B$. Suppose the boson can only hop within each sub-lattice, namely, we only consider next nearest neighbor hopping. The idea to realize the topological Mott insulator is the same as that we discussed in the double layer model. We first condense the boson and break the $U(1)$ symmetry. For instance, we can condense the $b_0$ boson by tuning the interactions [see eq.~(\ref{TrivialSPT2})].  
The second step is to restore the broken $U(1)$ symmetry by condensing the vortex field. A vortex becomes an anti-vortex under time reversal. The resultant state is a (trivial) Mott insulating phase, which can be realized by the following simple Hamiltonian,
\begin{eqnarray}\label{TrivialSPT2}
H&=&-\sum_{\langle\langle ij\rangle\rangle,\alpha,\beta}t_{ij}^{\alpha\beta} b_{\alpha,i}^\dag b_{\beta,j} +\sum_i [U(\hat N_i-1)^2+\mu \hat N_i\nonumber\\&&+D(S^z_i)^2],
\end{eqnarray}
where $\langle\langle\rangle\rangle$ means next nearest neighbor and $U>0, D>0$. Since $(S^z)^2=b_1^\dag b_1 +b_{-1}^\dag b_{-1}$, the $D$ term changes the chemical  potential of $b_1$ and $b_{-1}$ bosons.

To obtain the nontrivial Mott insulator, it is required that under time reversal the vortex reverses its sign and becomes anti-vortex. One way out is to redefine the vortex operator as $\tilde b'=\tilde bS^-$. Similar to previous discussion, we couple the bose current (carried by the vortex/anti-vortex) to the spin degrees of freedom (at the vortex/anti-vortex core).

Notice that we only considered next nearest neighbor hopping, so the two condensates on two different sub-lattices can be considered independent of each other. The vortex core of the condensate on A sub-lattice locates on B sub-lattice and vice versa (see Fig.\ref{fig:Int}b). The two condensates (on different two sub-lattices) couple to each other via the following interaction,
\begin{eqnarray} \label{int2}
H_{\rm int}=-\sum_i && g S_{i}^z(J_{i+y,i+x} + J_{i+x,i-y} + J_{i-y,i-x}\nonumber\\ && + J_{i-x,i+y}),
\end{eqnarray}
where $J_{ij}= \sum_{\alpha,\beta}(-ib_{\alpha,i}^\dag b_{\beta,j}+h.c.)$ 
is the current operator of the condensate, 
and $g> D$. Under above interaction, the vortex/anti-vortex tends to increase/decrease the spin momentum at its core and as a consequence vortex (anti-vortex) will change its sign under time reversal. According to our previous discussion of Chern-Simons effective field theory, this interaction will possibly stabilize the nontrivial SPT phase.

Similar idea can also be implemented in solid state systems.
In the supplemental material, we realize the BTI in a pure spin-$1$ systems in the limit $U\to+\infty$ where the `charge' degrees of freedom are frozen. The $U(1)$ symmetry of the system is defined by $U(1)=\{e^{i(3S_z^2-2I)\theta}\}$ that preserves spin nematic momentum instead of magnetic momentum, and the time reversal symmetry is defined as usual. The BTI phase realized this way corresponds to a topologically nontrivial spin nematic phase with gapless edge modes. 

\textit{Conclusion and discussion.} We constructed Bosonic topological
insulators protected by $U(1)\rtimes Z_2^T$ symmetry. We first let the boson
field condense and break the $U(1)$ symmetry. To gap out the Goldstone modes
and restore the symmetry, we condense the vortex/anti-vortex field. If the vortex/anti-vortex creation operator does not
reverse its sign under time reversal $T$, we obtain a trivial SPT phase. On the
other hand, if the vortex/anti-vortex creation operator flips the spin momentum at the vortex/anti-vortex core such that it reverses its sign under
$T$, we obtain a nontrivial SPT phase
--- the bosonic topological insulator after condensing the composite vortex. 
This mechanism might be realized in interacting superfluids/superconductors with spin-orbital coupling.  Based on above mechanism, we construct lattice models to realize BTI. Although our constructed Hamiltonians are still toy models, we mark an important step toward experimental realization of interacting bosonic SPT phases. For the purpose of detecting nontrivial BTI phase, we show that a $\pi$-flux in the bulk of BTI carries a Kramers doublet. 

Our method can be used to realize other SPT phases protected by different
symmetries, and might shed some light on experimental realizations. 
There are also some questions remaining open, for instance, the 
models we constructed have not been studied numerically, and we didn't discuss
how to identify the nontrivial SPT phases experimentally. These issues will be addressed 
in our further work.


We thank Peng Ye, Jia-Wei Mei, Juven Wang, Gang Chen and Xiong-Jun Liu for helpful
discussions. This research is supported in part by Perimeter Institute for
Theoretical Physics. Research at Perimeter Institute is supported by the
Government of Canada through Industry Canada and by the Province of Ontario
through the Ministry of Economic Development \& Innovation. ZXL thanks the
support from NSFC 11204149 and Tsinghua University Initiative Scientific
Research Program. XGW is also supported by NSF Grant No.  DMR-1005541,
NSFC 11274192, and the John Templeton Foundation.


\bibliography{wencross,all,publst}



\end{document}


\title{\large \bf Supplementary materials: Microscopic Realization of 2-Dimensional Bosonic Topological Insulators}

\author{Zheng-Xin Liu}
\affiliation{Institute for Advanced Study, Tsinghua University, Beijing, 100084, P. R. China}
\affiliation{Perimeter Institute for Theoretical Physics, Waterloo, Ontario, N2L 2Y5 Canada}
\author{Zheng-Cheng Gu}
\affiliation{Perimeter Institute for Theoretical Physics, Waterloo, Ontario, N2L 2Y5 Canada}
\author{Xiao-Gang Wen}
\affiliation{Institute for Advanced Study, Tsinghua University, Beijing, 100084, P. R. China}
\affiliation{Perimeter Institute for Theoretical Physics, Waterloo, Ontario, N2L 2Y5 Canada}
\affiliation{Department of Physics, Massachusetts Institute of Technology, Cambridge, Massachusetts 02139, USA}


\maketitle

\section{Double Chern-Simons theory of Higgs Mechanism}\label{CS_Higgs}

In the main text we use XY model to describe the low energy effective theory of the bose condensate. If the vortex of the boson condense, then $\theta$ in the XY model [see eqn.~(2) in the main text] is no longer a smooth function of space-time. We can introduce the singular part by replacing $\partial_\mu\theta$ by $\partial_\mu\theta+\tilde a_\mu$, where the field strength of gauge field $\tilde a_\mu$ correspond to the vortex current density $\tilde J^\mu={i\over 2\pi}\varepsilon^{\mu\nu\lambda}\partial_\nu \tilde a_\lambda$. The charge of gauge field $\tilde a_\mu$ is the number of vortices minus the number of anti-vortex and is quantized. This can be seen by integrating density $\tilde \rho$ in a closed space, which gives $\oint d^2x\tilde \rho={1\over2\pi}\oint d^2x(\partial_i\tilde a_j-\partial_j\tilde a_i)\in\mathbb Z$ owning to Gauss's law. The phase fluctuation of the vortex condensate can also be described by an XY model, which is equivalent to the Maxwell term of the gauge field $\tilde a_\mu$. Now 
the XY model becomes a Higgs Lagrangian
\begin{eqnarray}\label{XY2}
\mathcal{L}_{\rm Higgs}={1\over2}[(\partial_\mu \theta + \tilde a_\mu)^2 + {1\over4\pi^2}\tilde f_{\mu\nu}\tilde f^{\mu\nu}],
\end{eqnarray}
where $\tilde f_{\mu\nu}=\partial_\mu \tilde a_\nu-\partial_\nu \tilde a_\mu$ and we have normalized with $v=1, \chi=1$.

In the path integral
\begin{eqnarray}
Z=\int D\theta D\tilde a_\mu e^{-\int \mathcal{L}_{\rm Higgs}d^2xd\tau },
\end{eqnarray}

we can introduce a Hubbard-Stratonovich field $j_\mu$ to decouple the quadratic term as
\begin{eqnarray}
Z&=&\int D\theta D\tilde a_\mu Dj_\mu e^{-S },\nonumber\\
S&=&\int d^2xd\tau {1\over2}[(j_\mu)^2+2i\theta\partial_\mu j^\mu - 2ij^\mu \tilde a_\mu + {1\over4\pi^2}\tilde f_{\mu\nu}\tilde f^{\mu\nu}].\nonumber
\end{eqnarray}
The classical equation of motion for $j_\mu$ gives a gauge invariant current $j_\mu= i(\partial_\mu\theta + \tilde a_\mu)$. Integrating out the $\theta$ field results in a constraint $\partial_\mu j^\mu=0$. From this constraint, we can write $j^\mu={1\over2\pi}\varepsilon^{\mu\nu\lambda}\partial_\nu a_\lambda$. The charge of $a_\mu$ is equal to the boson number and is quantized. With these results, the path integral becomes
\begin{eqnarray}
Z&=&\int Da_\mu D\tilde a_\mu e^{-\int d^2xd\tau \mathcal L_{\rm CS} },\nonumber\\
\mathcal L_{\rm CS}&=&-{i\over2\pi}\varepsilon^{\mu\nu\lambda}\tilde a_\mu\partial_\nu a_\lambda+{1\over8\pi^2}[\tilde f_{\mu\nu}\tilde f^{\mu\nu}+f_{\mu\nu}f^{\mu\nu}],
\end{eqnarray}
where $f_{\mu\nu}=\partial_\mu a_\nu-\partial_\nu a_\mu$. Thus we finish the derivation of eqn.(3) in the main text.

\section{Realization of $U(1)\rtimes Z_2^T$ SPT phases in pure spin systems}\label{PureSpin}
Here we consider a spin-1 system with symmetry group  $U(1)\rtimes Z_2^T$. Here the $U(1)$ symmetry is defined as $U(1)=\{U_\theta|U_\theta=e^{i\Lambda\theta}, \Lambda=(3S_z^2-2I)=\mathrm {diag}(1,-2,1)\}$, which preserves the spin nematic order $\sum_iS_{zi}^2$. The generator $\Lambda$ is traceless because the `charge' degree of freedom is frozen. $Z_2^T$ is defined as $Z_2^T=\{I,T\}$, where $T=e^{iS_y\pi}K$ and $TU_\theta=U_{-\theta}T$. Notice that this $U(1)$ group is no longer a subgroup of $SU(2)$ spin rotation, instead, it is a subgroup of a larger group $SU(3)$. The eight generators (namely the Gell-Mann matrices) of the $SU(3)$ group can be written in forms of $S=1$ spin operators,
\begin{eqnarray}\label{GellMann}
\lambda_1&=&{1\over\sqrt2}(S_x+S_{xz}),\ \ \lambda_2={1\over\sqrt2}(S_y+S_{yz}),\nonumber\\
\lambda_4&=&S_x^2-S_y^2,\ \ \lambda_5=S_{xy},\nonumber\\
\lambda_6&=&{1\over\sqrt2}(S_x-S_{xz}),\ \ \lambda_7={1\over\sqrt2}(S_y-S_{yz}),\nonumber\\
\lambda_3&=&{1\over2}S_z+{3\over2}S_z^2-I,\ \  \lambda_8={1\over\sqrt3}({3\over2}S_z-{3\over2}S_z^2+I).
\end{eqnarray}
Here we have defined $S_{mn}=S_mS_n+S_nS_m,\ m,n=x,y,z$, and $I$ is the 3 by 3 identity matrix. The generators satisfy $\mathrm{Tr}(\lambda_a\lambda_b)=2\delta_{ab}$ and $[\lambda_a, \lambda_b]=if^{abc}\lambda_c$, where $f^{abc}$ are the completely antisymmetric structure factors with 
\begin{eqnarray}\label{Structure}
f^{123}=1,\ f^{147}=f^{165}=f^{246}=f^{257}=f^{345}=f^{376}={1\over2}, \ f^{458}=f^{678}={\sqrt3\over2}
\end{eqnarray}
and all others vanishing. The $SO(3)$ subgroup of $SU(3)$ generated by $S_x, S_y, S_z$ is commuting with the time reversal group $Z_2^T$.  The generator of the $U(1)$ symmetry is a linear combination $\Lambda={3\over2}\lambda_3-{\sqrt3\over2}\lambda_8$. From (\ref{GellMann}) and (\ref{Structure}), it is easy to verify that under the action of symmetry group $U(1)\rtimes Z_2^T$, the spin operators varies as the following: 
\begin{eqnarray}\label{symmetry}
&&U_\theta S_zU_\theta^{-1}=S_z,\ U_\theta S_{xy}U_\theta^{-1}=S_{xy},\ U_\theta S_m^2U_\theta^{-1}=S_m^2,\nonumber\\
&&U_\theta(S_x\pm iS_{yz})U_\theta^{-1}=e^{\pm i\theta}(S_x\pm iS_{yz}),\nonumber\\
&&U_\theta(S_{xz}\pm iS_y)U_\theta^{-1}=e^{\pm i\theta}(S_{xz}\pm iS_y),\nonumber\\
&&TS_xT^{-1}=-S_x, \ TS_yT^{-1}=-S_y,\ TS_zT^{-1}=-S_z,\ TS_{mn}T^{-1}=S_{mn}.
\end{eqnarray}

To construct the SPT phases, we first break the $U(1)$ symmetry. Notice that the $U(1)$ symmetry generated by $\Lambda={\rm diag}(1,-2,1)$ is not a spin rotation but is more like a `charge' conserving symmetry for the spin components $|1\rangle$ and $|-1\rangle$ (this can be more easily seen in the slave particle representation that will be discussed later). Breaking the $U(1)$ symmetry can be done by condensing the spin quantities $(S_{xz}+ iS_y)$ or  $(S_x+ iS_{yz})$, which can be considered as a bose field noted as $b_1$.  Restoring the symmetry requires to condense the vortex field of $b_1$, which is noted as $b_2$. Repeating the arguments given in the main text, we obtain a doubled Chern-Simons action
\begin{eqnarray}\label{doubleCS}
\mathcal L_{\mathrm{CS}}
&=&-{i\over4\pi}K_{IJ}\varepsilon^{\mu\nu\lambda}a_{I\mu}\partial_\nu a_{J\lambda} + \mathcal L_{\mathrm{Maxwell}},
\end{eqnarray}
where $J_1^\mu=\varepsilon^{\mu\nu\lambda}\partial_\nu a_2^\lambda$ and $J_2^\mu=\varepsilon^{\mu\nu\lambda}\partial_\nu a_1^\lambda$ are the bose current and vortex current, respectively. Here $b_I$  carry the charge of $a_I$ ($I=1,2$) gauge field. The boundary degrees of freedom satisfy the Kac-Moody algebra \begin{eqnarray}\label{KacMoody}
[\partial_x\phi_I,\partial_y\phi_J]=2\pi iK_{IJ}^{-1}\partial_x\delta(x-y),
\end{eqnarray}
where $a_I^\mu=\partial_\mu \phi_I$ and $b_I=e^{-i\phi_I}$ on the boundary. Under the action of the symmetry group, the $\phi$ fields vary as
\begin{eqnarray*}
&&U_\theta\phi_1U_\theta^{-1}=\phi_1+\theta, \  T\phi_1T^{-1}=-\phi_1,\\
&&U_\theta \phi_2U_\theta^{-1}=\phi_2, \  \ \ \ \ \ T\phi_2T^{-1}=\phi_2,
\end{eqnarray*}
or equivalently
\begin{eqnarray*}
&&U_\theta b_1U_\theta^{-1}=b_1e^{-i\theta}, \  Tb_1T^{-1}=b_1,\\
&&U_\theta b_2U_\theta^{-1}=b_2, \  \ \ \ \ \ \ Tb_2T^{-1}=b_2^\dag.
\end{eqnarray*}
This is a trivial SPT phase since the boundary can be gapped out by the perturbation term $G\cos\phi_2$, which fix $\phi_2=0$ without breaking any symmetry. So this state belongs to the trivial SPT phase.

\begin{figure}[t]
\centering
\includegraphics[width=2.in]{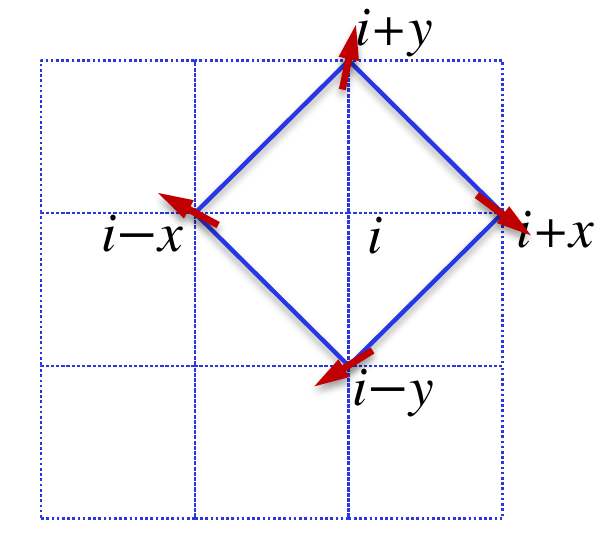}
\caption{(Color online)  Nontrivial SPT state on square lattice. The vortex (anti-vortex) core locates on lattice site.  The spin at the vortex (anti-vortex) core feel an effective field owning to the current carried by the vortex (anti-vortex).
} \label{fig:PureSpin}
\end{figure}

A direct product state with spin nematic order $\langle S_m^2\rangle\neq0$ or $\langle S_{xy}\rangle\neq0$ is a trivial SPT state since it is invariant under the symmetry group $U(1)\rtimes Z_2^T$. These trivial states can be easily realized in lattice models. For instance, the state with $\langle S_z^2\rangle=\langle S_y^2\rangle\neq0$  can be easily realized by the following model:
\begin{eqnarray}\label{trivial}
H=\sum_{\langle\langle ij\rangle\rangle}[J_x(S_i^xS_j^x+S_i^{yz}S_j^{yz})+J_y(S_i^yS_j^y+S_i^{xz}S_j^{xz})+J_zS_i^zS_j^z]+\sum_i[D_x(S_i^x)^2-D_y(S_i^y)^2],
\end{eqnarray}
where $\langle\langle \rangle\rangle$ means next nearest neighbor as discussed in the main text. If $D_y>D_x> J_{x,y,z}$, the ground state of above model can be approximately written as $\prod_i|x\rangle_i$, where $S_i^x|x\rangle_i=0$ (and $|y\rangle$ is defined similarly).

Now we construct the nontrivial SPT phase. To this end, we define a composite vortex operator $b_2'=b_2\hat O$, such that $b_2'$ is $U(1)$ neutral and reverse its sign under time reversal: $U_\theta b_2'U_\theta^{-1}=b_2',\ Tb_2'T^{-1}=-b_2'^\dag$. From (\ref{symmetry}), only $\hat O=S_z$ satisfy above conditions.  If we write $b_2'=e^{-i\phi_2'}$ on the boundary, then 
\[
U_\theta\phi_2'U_\theta^{-1}=\phi_2', \ T\phi_2'T^{-1}=\phi_2'+\pi.
\] 
If we condense the composite vortex $b_2'$ then we obtain a nontrivial SPT phase where 
the boundary cannot be gapped without breaking symmetry.  

Since $S_z|x\rangle=|y\rangle$, a vortex/anti-vortex will transform the spin state from $|x\rangle$ to $|y\rangle$. If the spin is already staying in state $|y\rangle$, a vortex/anti-vortex will transform it back to $|x\rangle$ because $S_z|y\rangle=|x\rangle$. So we need to design an interaction such that creating a vortex/anti-vortex will switch the spin state at the core between $|x\rangle$ and $|y\rangle$.  A possible interaction is (see Fig.\ref{fig:PureSpin})
\begin{eqnarray}
H_{\rm int} = \sum_i g (S_i^y)^2  (F_{i+y,i+x}+F_{i+x,i-y}+F_{i-y,i-x} + F_{i-x,i+y})^2,
\end{eqnarray}
where $g> D_y$ and $F_{i+u,i+v}={1\over2}[-i(S^{xz}+iS^y)_{i+u}(S^{xz}-iS^y)_{i+v}+h.c.]=S_{i+u}^yS_{i+v}^{xz}-S_{i+u}^{xz}S_{i+v}^y$ is the $U(1)$ `vorticity' operator for the two spins at sites $i+u$ and $i+v$ relative to site $i$. Here the interaction contains a square term of the vorticity 
because the vortex and anti-vortex creating operators act on the core spin in the same way.  Adding above interaction to the model (\ref{trivial}) may stabilize the nontrivial SPT phase.

The construction of SPT phases can be more easily understood in the slave particle representation of spins.  We introduce three components of (slave) boson field $d=(d_1,d_0,d_{-1}, )^T$ to represent the spin operator $S_m=d^\dag I_m d$ under the particle number constraint $d_1^\dag d_1+d_0^\dag d_0+d_{-1}^\dag d_{-1}=1$, where $I_m$ is the three-by-three matrix representation of spin operator $S_m$. The bosons transform in the following way under the symmetry action: 
\begin{eqnarray*}
&&U_\theta d_1U_\theta^{-1}=d_1e^{-i\theta},\ U_\theta d_{-1}U_\theta^{-1}=d_{-1}e^{-i\theta},\ U_\theta d_0U_\theta^{-1}=d_0,\\
&&Td_1T^{-1}=d_{-1}, \ Td_{-1}T^{-1}=d_{1}, \ Td_0T^{-1}=-d_0. 
\end{eqnarray*}
Notice that there is a $U(1)$ gauge symmetry for the boson field $d$ since the spin operator $S_m$ is invariant under $U(1)$ gauge transformation $d\to de^{i\varphi}$, $d^\dag\to d^\dag e^{-i\varphi}$. The $U(1)$ gauge fluctuations couple to the slave particles $d$ as internal $U(1)$ gauge field. However, this $U(1)$ gauge field will be Higgsed if the bosons condense. In our later discussion,  the internal $U(1)$ gauge field is always Higgsed by, for example, the $d_0$ condensate.

To construct the SPT phases, we first break the $U(1)$ symmetry. We condense the boson fields $b_1=(d_1-d_{-1})$ (if $d_0$ also condenses, then the spin quantity $S_{xz}+iS_y=\sqrt2b_1^\dag d_0$ mentioned previously will form a classical order). The second step is restoring the symmetry by condensing the vortex field of $b_1$, which is noted as $b_2$. The vortex condensate $b_2$ together with the internal $U(1)$ gauge field kills the phase coherence of the bose condensates. So finally $U(1)$ symmetry is restored but the density $b_1^\dag b_1$ is still finite. The bose field $b_1$ and the vortex field $b_2$ vary as the following under symmetry group:
\begin{eqnarray*}
&&U_\theta b_1U_\theta^{-1}=b_1e^{-i\theta}, \  Tb_1T^{-1}=-b_1,\\
&&U_\theta b_2U_\theta^{-1}=b_2, \  \ \ \ \ \ \ Tb_2T^{-1}=b_2^\dag.
\end{eqnarray*}
Here the minus sign in $Tb_1T^{-1}=-b_1$ can be removed by redefining $b_1\to ib_1$. The effective field theory is also a Chern-Simons action. Above state is a trivial state, and the nontrivial one can be obtained by defining a composite vortex operator $b_2'=b_2S_z$. The construction of the lattice Hamiltonian of the nontrivial phase is the same as our previous discussion. The bose field $b_1$ is defined as $b_1=(d_1-d_{-1})$ because the onsite anisotropy term $\sum_iD(S_i^x)^2$ in eq.(\ref{trivial}) favors finite density $b_1^\dag b_1$.
